# The coexistence of magnetism and ferroelectricity in 3d transition metal doped SnTe monolayer

Yanyu Liu,† Wei Zhou, ‡ Gang Tang,† Chao Yang,† Xueyun Wang,† and Jiawang Hong†*

†School of Aerospace Engineering, Beijing Institute of Technology, Beijing 100081, China

‡Department of Applied Physics, Faculty of Science, Tianjin University, Tianjin 300072, P.R. China

## ABSTRACT

The realization of multiferroicity in 2D nanomaterials is crucially important for designing advanced nanoelectronic devices such as non-volatile multistate data storage. In this work, the coexistence of ferromagnetism and ferroelectricity is reported in monolayer SnTe system by transition metal (TM) doping. Based on first-principles calculations, the spontaneous spin polarization could be realized by TM doping in ferroelectric SnTe monolayer. In addition to in-plane ferroelectric polarization, the out-of-plane ferroelectric polarization emerges in Mn (Fe)-doped SnTe monolayer due to the internal displacement of TM from the surface. Interestingly, the crystalline field centered on TM and interaction between the dopant and Te gradually enhanced with the increment of atomic number of doping elements, which explains why the formation energy decreases. The realization of multiferroics in SnTe monolayer could provide theoretical guidance for experimental preparation of low-dimensional multiferroic materials.

* Corresponding author. E-mail: hongjw@bit.edu.cn (J Hong); Tel: +86 010 68915917

# INTRODUCTION

Multiferroic materials, in which ferroelectricity coexists with magnetism, have attracted growing attentions, due to their fascinating physical properties and novel functionalities.[1-4] However, in 2-dimensional (2D) materials, the enhanced depolarization field usually destroys the ferroelectricity below the critical thicknesses.[5, 6] Meanwhile, according to the Mermin-Wagner theorem, the long-range magnetic order is strongly suppressed by thermal fluctuations.[7, 8] Therefore, it is still a great challenge to retain the ferroelectricity and long-range magnetic ordering stable in thin films at room temperature to date.[9] Despite considerable efforts in searching for 2D ferroelectric (FE) and ferromagnetic (FM) materials with experimental and theoretical works, only a handful of 2D FE (e.g. SnTe[10] and $CuInP_2S_6$[11], $MX_2$ (M=Mo, W; X=S, Se, Te)[12-15], etc) and FM (e.g. $Cr_2Ge_2Te_6$[8], $CrI_3$[16] and $Fe_3GeTe_2$[17]) materials are synthesized in experiment. In addition, some theoretical work foretold that FM and FE could coexist in a few materials[18-20] such as layered transition-metal halide systems[18], 2D hyperferroelectric metals[19], but it still needs more efforts for the practical application of multiferroic materials with ferromagnetism and out-of-plane ferroelectricity.

One of promising ways to realize the coexistence of FE and FM is to decorate them with 3d transition-metal (TM) in FE materials, because TM doping is one of common methods to induce magnetism in nonmagnetic 2D materials[21-23]. Here, we choose FE SnTe as host material and introduce Mn, Fe and Co as dopants. The simple structure and chemical compositions of SnTe monolayer is beneficial to investigate its microscopic mechanism for multiferroic properties. Moreover, in-plane FE has been observed and manipulated in SnTe film[10], and theoretically, its ferroelectricity exhibits non-monotonic thickness dependence[24], which beyond the conventional scaling law[18, 25]. Recently, Wang et al. succeeded in fabricating high-quality and Cr-doped SnTe

(111) thin films, exhibiting ferromagnetism[26]. As usual, the FE properties is highly sensitive to structural properties. Therefore, after TM doping, it needs further investigation to confirm whether the experimentally observed robust ferroelectricity are retained or not. Here, we systematically investigate the influence of distortion on ferroelectric and magnetic properties of FE SnTe monolayer by introducing different TM dopants. The introduction of TM (Fe and Mn) surprisingly induces out-of-FE plane polarization and spontaneous spin polarization, keeping the large in-plane FE polarization. Furthermore, the Fe (Mn)-doped SnTe monolayer simultaneously holds electric polarization and spin polarization, especially the appearance of the out-of-plane polarization, which could pave the way for 2D materials in potential applications.

## COMPUTATIONAL METHODS

All calculations are carried out within the framework of density functional theory methods implemented in the Vienna ab initio Simulation Package (VASP) code[24-25]. The exchange–correlation effect described within generalized gradient approximation (GGA) with the Perdew−Burke-Ernzerhof (PBE) functional[26], together with the projector augmented wave method[27], are adopted in DFT calculations. The doped model is a 3×3×1 periodic supercell with a 3×3×1 Monkhorst–Pack k-point mesh is used and the kinetic energy cutoff of 450 eV is set. In order to avoid the interlayer interaction due to the periodic boundary condition, a vacuum layer of 20 Å is adopted. All the doped systems are fully relaxed until the force on each atom is less than 0.01 eV/Å, and the convergence criteria for energy of $1 \times 10^{-5}$ eV is satisfied. In addition, a convergence test with a larger energy cutoff and denser k-points yields nearly the same results. In order to accurately describing the electronic properties of 3d TM, the hybrid functional (HSE06[28-29]) is employed in this work. The Berry-phase method[30] is performed to evaluate the electrical polarization.

## RESULTS AND DISCUSSION

## Pristine SnTe Monolayer

Our previous work[24] reveals that the ferroelectric structure with [110]-mode which is in consistent with the experimental observation[10]. The corresponding monolayer is displayed in Figure 1(a), which indicates the large in-plane FE polarization in SnTe monolayer results from the relative displacement of the Sn and Te atoms along [110] direction. The calculated lattice constants and ferroelectric polarization of FE SnTe monolayer have been summarized in Table 1, together with the reported theoretical and experimental data. The ferroelectric SnTe monolayer with a hinge-like structure has in-plane spontaneous polarizations $P_s$=25.83 μC/cm$^2$.

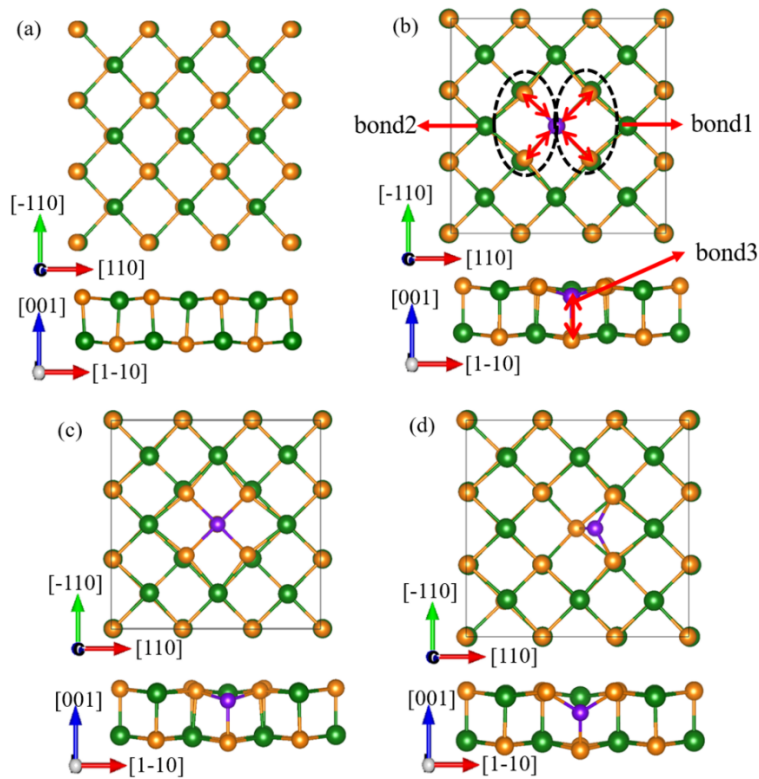

**Figure 1.** The geometric structure: (a) the pristine SnTe, (b)Sn$_{1-x}$Mn$_x$Te monolayer with different bonds, (c) Sn$_{1-x}$Fe$_x$Te and(d) Sn$_{1-x}$Co$_x$Te. The green, orange, purple spheres represent Sn, Te, and TM atoms, respectively.

**Table 1** Calculated lattice constants *a* and *b* (in Å) with spontaneous polarization $P_s$ (in μC/cm$^2$).

|  | *a* | *B* | $P_s$ |
|---|---|---|---|
| Experiment[10] | 4.58±0.05 | 4.44±0.05 | - |
| Theoretical[31] | 4.67 | 4.58 | 19.4/22.0[32] |
| This work | 4.67 | 4.55 | 25.83 |

## Structural properties

Next, TM-doped SnTe monolayer with one Sn atom replaced by a TM dopant per supercell (Sn$_{1-x}$TM$_x$Te) is taken into account, as illustrated in Figure 1(b). To verify the stability of these defective systems, the defect formation energy is estimated from the following definition:

$$E_f = E_{doped} - E_{\text{pure}} - \mu_{\text{TM}} + \mu_{\text{Sn}} \qquad (1)$$

where $E_{dope}$ and $E_{pure}$ are the total energy of the supercell with one Sn$_{\text{TM}}$ defect (one Sn atom substituted by a TM atom) and the perfect supercell, respectively. The $\mu_{\text{TM}}$ and $\mu_{\text{Sn}}$ are, respectively, the chemical potential of TM and Sn, which are obtained from the constituent elements in their standard states. According to the equation (1), the formation energy for each doped system has been listed in Table 2. The formation energies are negative, indicating these defective systems are energetically favourable. This should be attributed to the absence of the outward atoms, yielding the strain induced by the dopant easily release. Furthermore, it can be found that the formation energy decreases with the increase of the atomic number of the dopant.

In order to reveal the mechanism, we investigate the local structure surrounding the dopants, such as the bond length of the TM with its nearest Te. Based on the crystal symmetry, TM atom forms three different bonds with its nearest Te, bond1, bond2 and bond3, as illustrated in Figure 1(b). The bond lengths for different dopant are summarized in Table 2. From Table 2, it is observed that as the size of dopant increase, the bond1 and bond3 gradually shorten, while the bond2 visually shortens and then lengthens. In fact, for the Co-doped system, the bond2 between TM and Te increases to more than 4 Å which suggests the bond has been broken. It results in the crystalline field centred on TM transforms to plane triangle from the broken octahedron, in which one of the apical Te atom is missing. With the atomic number of doping elements increasing, the bond related to crystalline field decrease. It means that the crystalline field centred on TM gradually enhanced with the increment of atomic number of doping elements. As a consequence, the interaction between the TM and Te atoms is strengthened with the increment of atomic number of doping elements, which coincides with the feature of electronic structures of $Sn_{1-x}TM_xTe$ monolayers in the following. Figure 1(b)-(d) visually present the variance of lattice distortion induced by the TM doping.

**Table 2** Calculated bond length (in Å) between TM and its nearest Te include three different kind bonds. $E_f$ represents the defect formation energy (in eV) for $Sn_{1-x}TM_xTe$ monolayer.

|      | $E_f$   | Bond1 | Bond2 | Bond3 |
|------|---------|-------|-------|-------|
| Pure | -       | 3.133 | 3.363 | 2.926 |
| Mn   | -1.921  | 2.942 | 2.957 | 2.751 |
| Fe   | -1.663  | 2.824 | 2.956 | 2.651 |
| Co   | -1.472  | 2.598 | 4.183 | 2.508 |

## Magnetic properties

As discussed above, the transition of the crystalline field centred on the dopants is enhanced with the increase of the atomic number of the dopant. It means that the interaction between the dopant with the nearest neighbour Te atoms is enhanced, which agrees well with the electronic structure of the doped systems. The electronic structures of the doped systems are depicted in Figure 2 where we can tell that the hybridization of the d orbital of TM with the p orbital of Te is strengthened with the increase of the atomic number of the dopant. Moreover, the hybridization of Co d orbital and Te orbital is much stronger than that of Mn (Fe), especially in the energy range -6 to 0 eV, as depicted in Figure 2. The strong interaction in the former lifts the valence band maximum. As a result, the transformation occurs from semiconductor to metal for the Sn1-xCoxTe monolayer (see Figure 2). The crystalline field of the broken octahedron and the plane triangle is presented in Figure 2(d).

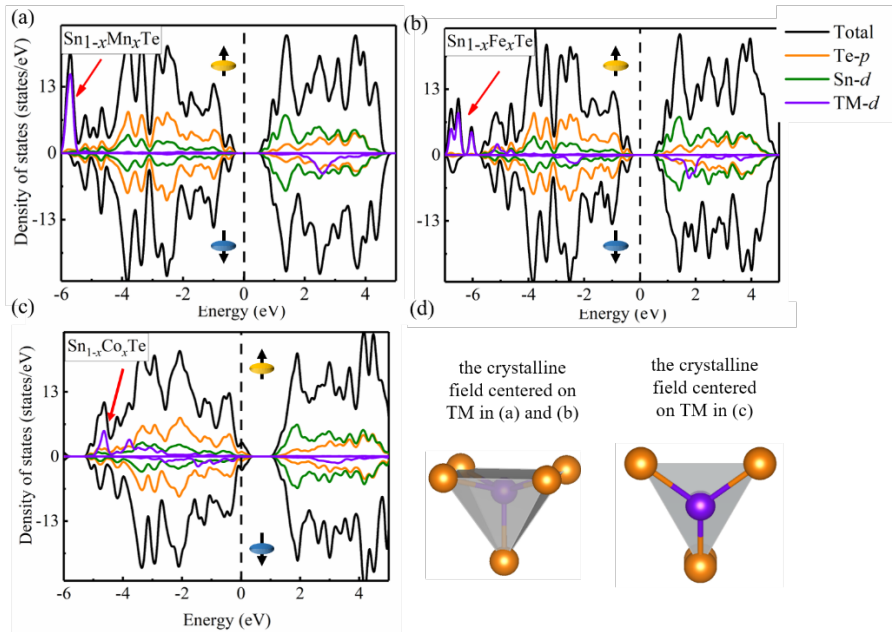

**Figure 2.** Spin resolved total and projected DOS for the $Sn_{1-x}M_xTe$ monolayers:(a) $Sn_{1-x}Mn_xTe$ (b) $Sn_{1-x}Fe_xTe$ (c) $Sn_{1-x}Co_xTe$, (d) the crystalline field in $Sn_{1-x}TM_xTe$. The spin-up and spin-down are marked by the yellow and blue ball with arrow, respectively.

From the Figure 2, it also can be found the spin-polarized character for the $Sn_{1-x}TM_xTe$ monolayers are induced from the asymmetry of the majority- and minority-spin channels. The detailed magnetic moments in $Sn_{1-x}TM_xTe$ are listed in Table 3. The total magnetic moments of $Sn_{1-x}TM_xTe$ are 5.00, 4.00 and 3.00 $\mu_B$ for Mn-, Fe- and Co-doped systems, respectively. As shown in Figure 2, the total and projected density of states (DOS) implies that the magnetic moment mainly originates from the d orbital of TM atoms. Furthermore, we note that most of spin-polarized electrons in the $Sn_{1-x}TM_xTe$ monolayer locate at the dopant site, resulting in a large magnetic moment of 4.20 $\mu_B$/Mn, 3.54 $\mu_B$/Fe and 2.37 $\mu_B$/Co. The strong magnetization of the TM ions can also be well confirmed by the spatial spin density distributions (see Figure 3).

**Table 3** Magnetic moment in $Sn_{1-x}TM_xTe$ monolayers. $\mu_{tot}$, $\mu_{TM}$ and $\mu_d$ indicate the magnetic moment of the supercell, the TM atom, the $d$ orbital of TM, respectively.

|  | Mn | Fe | Co |
|---|---|---|---|
| $\mu_{tot}$ | 5.00 | 4.00 | 3.00 |
| $\mu_{TM}$ | 4.20 | 3.54 | 2.37 |
| $\mu_d$ | 4.13 | 3.48 | 2.32 |

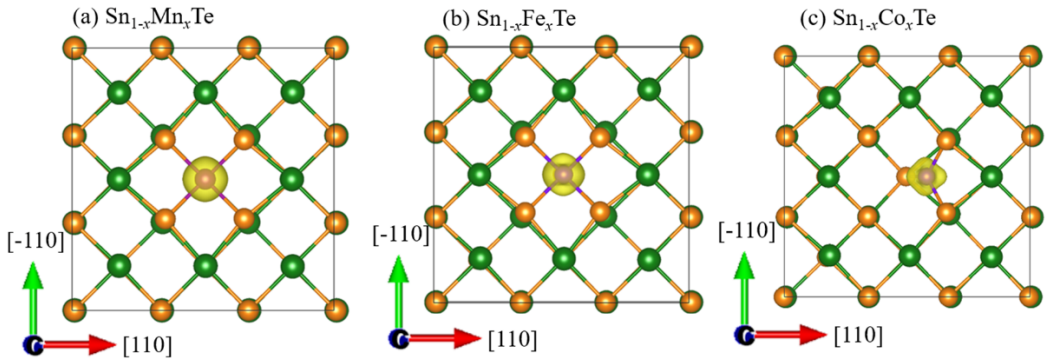

**Figure 3.** The spatial distribution of the spin density and the polyhedral crystal field for the $Sn_{1-x}TM_xTe$ monolayers: (a) $Sn_{1-x}Mn_xTe$ (b) $Sn_{1-x}Fe_xTe$ (c) $Sn_{1-x}Co_xTe$. The isosurface is 0.01 $\mu_B$/Å$^3$. The green, orange, purple spheres stand for Sn, Te, and TM atoms, respectively.

Apparently, the magnetism ultimately depends on the electronic localization and spin polarization of these 3d TM ions. When one TM dopant donates two s-orbital electrons to bond with surrounding Te atoms, it results in a $d^5$, $3d^6$, and $3d^7$ configuration for $Mn^{2+}$, $Fe^{2+}$, and $Co^{2+}$, respectively. For Mn-doped monolayer, five d-electrons fully occupy the majority-spin states. Then, as the d electrons increasing of Fe and Co, the minority-spin states will be gradually occupied. Therefore, there are five, four, three and one unpaired d electron for the $Sn_{1-x}Mn_xTe$, $Sn_{1-x}Fe_xTe$ and $Sn_{1-x}Co_xTe$ monolayers, respectively. The relationship between magnetic moment and the atomic number of TM in $Sn_{1-x}TM_xTe$ e monolayer could be well understood by the spin resolved electron occupied configuration in Figure 4.

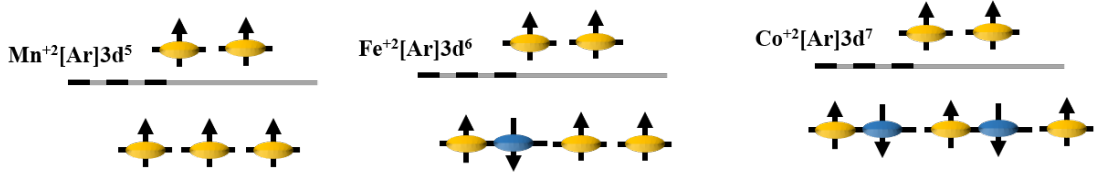

**Figure 4.** The spin-resolved electron occupied configurations of 3d states of TMs in polyhedral crystal field. The spin-up and spin-down are marked by the yellow and blue ball with arrow, respectively.

## Ferroelectric properties

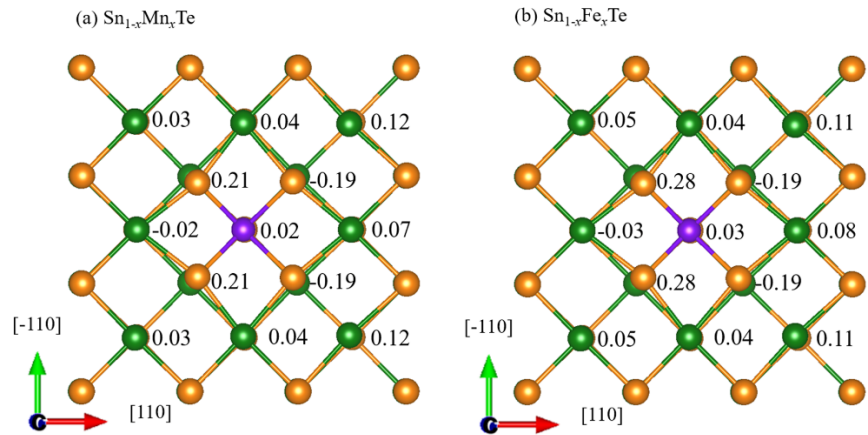

**Figure 5.** The referred positions of relative displacement of TM/Sn with Te for the $Sn_{1-x}Mn_xTe$ (a) and $Sn_{1-x}Fe_xTe$ (b) monolayers. The displacement in Å of Sn/TM and Te atoms in different positions. The green, orange, purple spheres stand for Sn, Te, and TM atoms, respectively.

The FE properties of SnTe monolayer come from the relative displacement of sub-lattice between Sn and Te atoms. As usual, the FE properties is very sensitive to the structure distortion. As stated above, the incorporation of TM induces large distortion in $Sn_{1-x}TM_xTe$ monolayers. Therefore, to verify the coexistence of the ferromagnetism and ferroelectricity, the dependence of Mn and Fe doping on the FE of SnTe is investigated here (Co doped-SnTe are not considered here due to their metallicity). For the pristine of SnTe monolayer, the relative displacement of sub-lattice between Sn and Te is 0.16 Å. For the doping systems, the Te atoms move towards to the nearest neighbour TM/Sn atoms, as marked as in Figure 5. As shown in Figure 5, the farther away from the doping position is, the larger the relative displacement is. As a consequence, the global relative displacement of cation-anion atoms is reduced for the $Sn_{1-x}Mn_xTe$ and $Sn_{1-x}Fe_xTe$ monolayers. And, the calculated in-plane spontaneous polarizations are 9.56 and 12.68 $\mu C/cm^2$ for $Sn_{1-x}Mn_xTe$ and $Sn_{1-x}Fe_xTe$ monolayers, respectively. Apparently, the decrease of the spontaneous polarizations agrees well with the global relative displacement of cation-anion atoms for the $Sn_{1-x}Mn_xTe$ and $Sn_{1-x}Fe_xTe$ monolayers.

Interestingly, we find that the TM doping also induce the out-of-plane polarization in SnTe monolayer. As presented in Figure 1(b) and (c), it can be observed that the Mn and Fe atom move inward from the surface which results in the out-of-plane spontaneous polarization. Obviously, the structure distortion highly affects the FE properties of the doped systems. The out-of-plane spontaneous polarizations are 2.13 and 2.21 $\mu C/cm^2$ in $Sn_{1-x}Mn_xTe$ and $Sn_{1-x}Fe_xTe$ monolayers, respectively.

## CONCLUSION

In summary, we proposed a promising alternative way to realize the coexistence of FE and FM by doping Mn and Fe into SnTe monolayer. Our work revealed that the crystalline field centred on TM transforms to plane triangle from the octahedron in Co doped monolayer while it keeps the octahedron filed in Mn and Fe doped monolayers. The change of the crystalline field has no effect on the high spin configuration of for $Mn^{2+}$ and $Fe^{2+}$ and $Co^{2+}$. Therefore, the magnetic moment of $Sn_{1-x}TM_xTe$ is 5.00, 4.00 and 3.00 $\mu_B$ for Mn-, Fe- and Co-doped systems, respectively. The less distortion in doped systems weakens the original in-plane polarization, but simultaneously introduce the out-of-plane polarization in $Sn_{1-x}Mn_xTe$ and $Sn_{1-x}Fe_xTe$ monolayers. Our findings of co-existence of ferroelectricity and magnetism in 2D $Sn_{1-x}Mn_xTe$ and $Sn_{1-x}Fe_xTe$ monolayers could provide theoretical guidance for designing low-dimensional multiferroic materials.

AUTHOR INFORMATION

**Corresponding Author**

*E-mail: hongjw@bit.edu.cn

**Notes**

The authors declare no competing financial interest.

**ACKNOWLEDGMENT**

The work at Beijing Institute of Technology is supported by National Natural Science Foundation of China with Grant Nos. 11572040, 11604011, 11804023, 51672007, and the China Postdoctoral Science Foundation with Grant No. 2018M641205. Theoretical calculations were performed using resources of the National Supercomputer Centre in Guangzhou, which was supported by Special Program for Applied Research on Super Computation of the NSFC-Guangdong Joint Fund (the second phase) under Grant No. U1501501.